\documentclass[submission,copyright,creativecommons]{eptcs}
\providecommand{\event}{Wivace 2013} 
\usepackage{breakurl}                
\usepackage{graphicx}                

\title{Evolution and development of complex computational systems using the paradigm of metabolic computing \\
   in Epigenetic Tracking}
\author{Alessandro Fontana, Borys Wr\'obel
\institute{Evolving Systems Laboratory, Faculty of Biology, Adam Mickiewicz University, Pozna\'n, Poland}
\email{fontana@evosys.org, wrobel@evosys.org}
}
\def\titlerunning{Evolution and development of complex computational systems}
\def\authorrunning{Alessandro Fontana, Borys Wr\'obel}
\begin{document}
\maketitle

\clubpenalty=10000 
\widowpenalty=10000 

\begin{abstract}
Epigenetic Tracking (ET) is an Artificial Embryology system which allows for the evolution and development of large complex structures built from artificial cells. In terms of the number of cells, the complexity of the bodies generated with ET is comparable with the complexity of biological organisms. We have previously used ET to simulate the growth of multicellular bodies with arbitrary 3-dimensional shapes which perform computation using the paradigm of ``metabolic computing''. In this paper we investigate the memory capacity of such computational structures and analyse the trade-off between shape and computation. We now plan to build on these foundations to create a biologically-inspired model in which the encoding of the phenotype is efficient (in terms of the compactness of the genome) and evolvable in tasks involving non-trivial computation, robust to damage and capable of self-maintenance and self-repair.
\end{abstract}

\section{Introduction}

Artificial Embryology (AE) is a field of Artificial Life which aims to build bio-inspired complex structures (usually consisting of small units, cells). There are two broad categories of AE approaches: grammatical and chemical. The grammatical approach uses grammatical rewrite rules to construct the description of the final structure step-by-step. An example is given by L-systems, first introduced by Lindenmayer \cite{Lindenmayer68} to model the growth of plants, but there is a number of other examples, based on context-free or context-sensitive grammars, instruction trees or directed graphs, e.g. \cite{Degaris99, Gruau96}. The chemical models, in comparison with grammatical models, are directly inspired by biological mechanisms of control at the sub-cellular level (in particular, gene regulatory networks) and mechanisms of communication between cells (e.g., through the diffusion of regulatory chemical substances). One such model (based on reaction-diffusion equations) was proposed already by Alan Turing \cite{Turing52} to explain the striped patterns observed in Nature. Recent examples of chemical models include \cite{Joachimczak08, Miller03}.

Epigenetic Tracking (ET), first described in \cite{Fontana08}, is a model of evolution of multicellular embryogenesis which strives to combine the advantages of the grammatical models, especially in terms of computational efficiency, and the cellular models, in terms of biological plausibility. Both efficiency and plausibility are relevant from the point of view of possible engineering applications of AE models: the hope of the field is that inspiration from biology will allow to build systems which are robust to damage, and possibly also self-maintaining and self-repairing. But potential applications are only one aspect of AE, another is the understanding-by-building attitude: the hope that by building biologically-inspired artificial systems we can explain the general rules which govern biological life.

In our previous work, we have demonstrated that ET is able to generate arbitrary 3-dimensional (3D) multicellular structures starting from a single cell, and that the model has interesting biological implications \cite{Fontana09}. The present work is based on a version of the model introduced in \cite{Fontana10}, which includes a biologically-inspired form of computation termed ``metabolic computation'' \cite{Fontana10}. How does this model differ from other bio-inspired models of computation? In many ways, the use of artificial gene regulatory networks and artificial neural networks for computation share the same spirit, even if the biological inspiration is different: artificial gene networks are inspired by sub-cellular processes, while artificial neural networks by a form of inter-cellular communication. Metabolic computation draws the inspiration from yet another biological process, in which chemical substances are processed by specialised cells in a multicellular organ, and involves modelling at both intra- and inter-cellular level.

In metabolic computation, a cell at a specific pre-specified location in the multicellular body is provided by an input (a vector of concentration values for a number of chemical substances), and the whole structure can then generate an output, read out from a cell at another pre-specified location (the output if also a vector of numbers: of concentration values for the same substances). Artificial evolution can be used to store a number of input-output pairs in the structure. This conversion of input to output can be viewed as a form of computation or a form of memory: the correspondence between inputs and outputs is stored in the multicellular structure obtained through evolution and development. In this work we ask what is the memory capacity of a structure obtained with ET. 

\section{Epigenetic Tracking as a model of multicellular development}
\label{sec:2}

\begin{figure}[t] \begin{center}
{\fboxrule=0.0mm\fboxsep=0mm\fbox{\includegraphics[width=16.00cm]{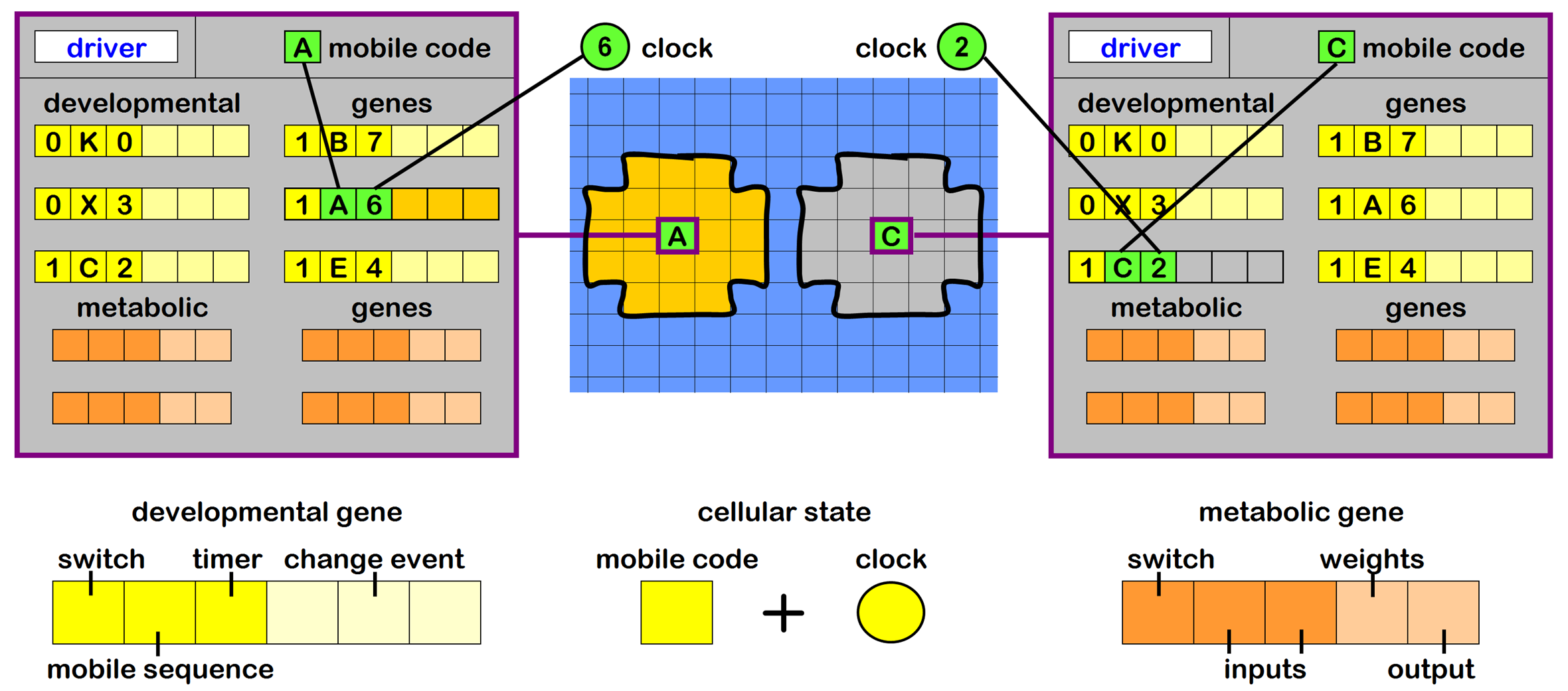}}}
\caption{Driver cells and change events in Epigenetic Tracking. On the left side, a driver cell with mobile code A triggers a proliferation event. On the right side, a driver cell with mobile code C triggers an apoptosis event.}
\label{chevents}
\end{center} \end{figure}

Multicellular bodies in ET consist of cube-shaped cells on a 3D grid. The growth of the embryo starts from a single cell placed in the middle of the grid, and unfolds in a pre-set number of discrete \emph{developmental stages} counted by a \emph{global clock}. Cells belong to two categories: \emph{normal} and \emph{driver}. All cells have an associated genome, which is a list of a genes. Driver cells have an additional variable called \emph{mobile code}, which can be interpreted as the set of cellular regulatory factors, and allows cells to behave differently despite sharing the same genome.

Two kinds of genes exist: \emph{developmental genes}, which specify development, and \emph{metabolic genes}, responsible for metabolic computation. Developmental genes are composed of a left part and of a right part. The left part contains a field called \emph{switch} which specifies if the gene is active or inactive, a field called \emph{mobile sequence}, and a field called \emph{timer}. At each developmental stage, for each driver cell, and for all developmental genes, the mobile code of a cell is compared with the mobile sequence of a gene and the timer is compared with the clock. If both match, a \emph{change event}, encoded in the gene's right part, occurs. One field in the right part determines if the driver cell in which the match occurs will cause a proliferation or an apoptosis (removal of cells) in the surrounding volume (Fig.~\ref{chevents}). Another field specifies the shape of the local structure created during a proliferation event (an ellipsoid of variable elongation) or the shape of the volume in which cells are removed. The third field of the right part determines the final differentiation state of the normal cells created in the proliferation, represented by their colour. 

In case of proliferation, both normal cells and driver cells are generated. Normal cells fill the local volume, driver cells (much fewer in number) are evenly distributed in the volume. Each new driver cell is assigned a new and unique mobile code value. This assignment is based on the mobile code of the driver that orchestrated the change event, and drivers created in this way form lineages. This is inspired by so-called ``mosaic'' developmental mechanisms \cite{Wolpert10}. If some developmental gene in the genome has a mobile sequence that matches the code of a newly generated driver cell, this cell can become the centre of another proliferation or apoptosis event at a subsequent stage.

\begin{figure}[t] \begin{center}
{\fboxrule=0.0mm\fboxsep=0mm\fbox{\includegraphics[width=11.50cm]{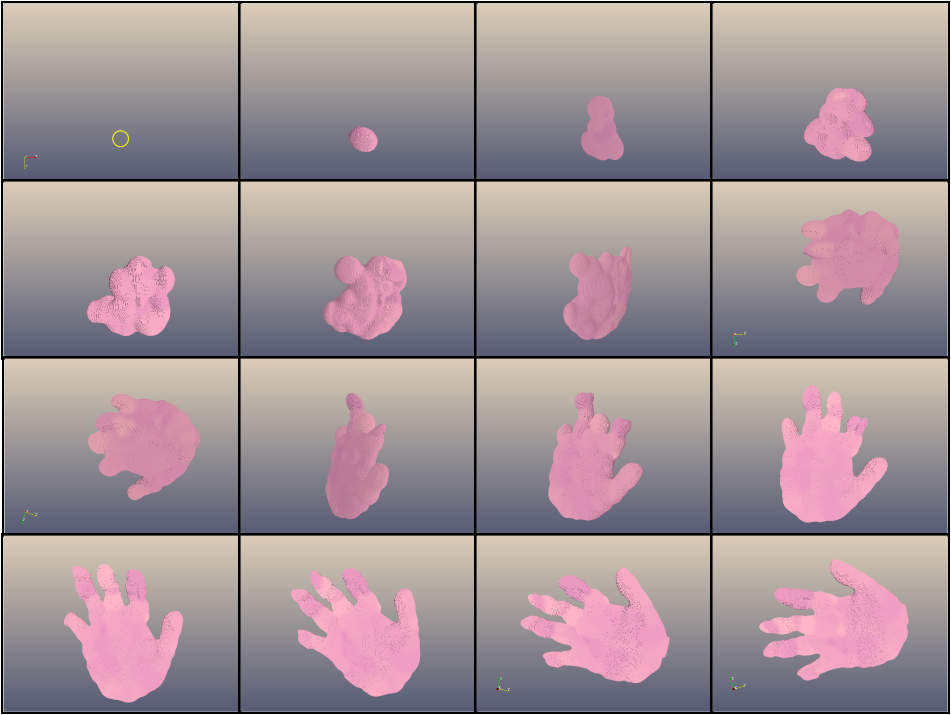}}}
\caption{Developmental sequence of a hand-shaped 3D structure obtained with ET, composed of 3.3 million cells.}
\label{hand}
\end{center} \end{figure}

Two key features distinguish ET from other models used in AE. First, in ET development is orchestrated through a small subset of cells (drivers), fewer than normal cells by several orders of magnitude. Second, in a single change event (proliferation or apoptosis) many cells can be created or deleted at once (instead of one, as in most other AE models). These features draw inspiration from the hierarchical aspects of biological development (e.g, the presence of organisers or stem cells \cite{Wolpert10}). It is this combination of elements which makes the genotype-phenotype relation in ET extremely efficient: compact genomes can specify very large multicellular structures, comparable in size with biological multicellular organisms (for example, 80 genes are enough to control the development of the hand-like structure with 3.3 million cells in Fig.~\ref{hand}). 

\begin{figure}[t] \begin{center}
\includegraphics[height=11.50cm]{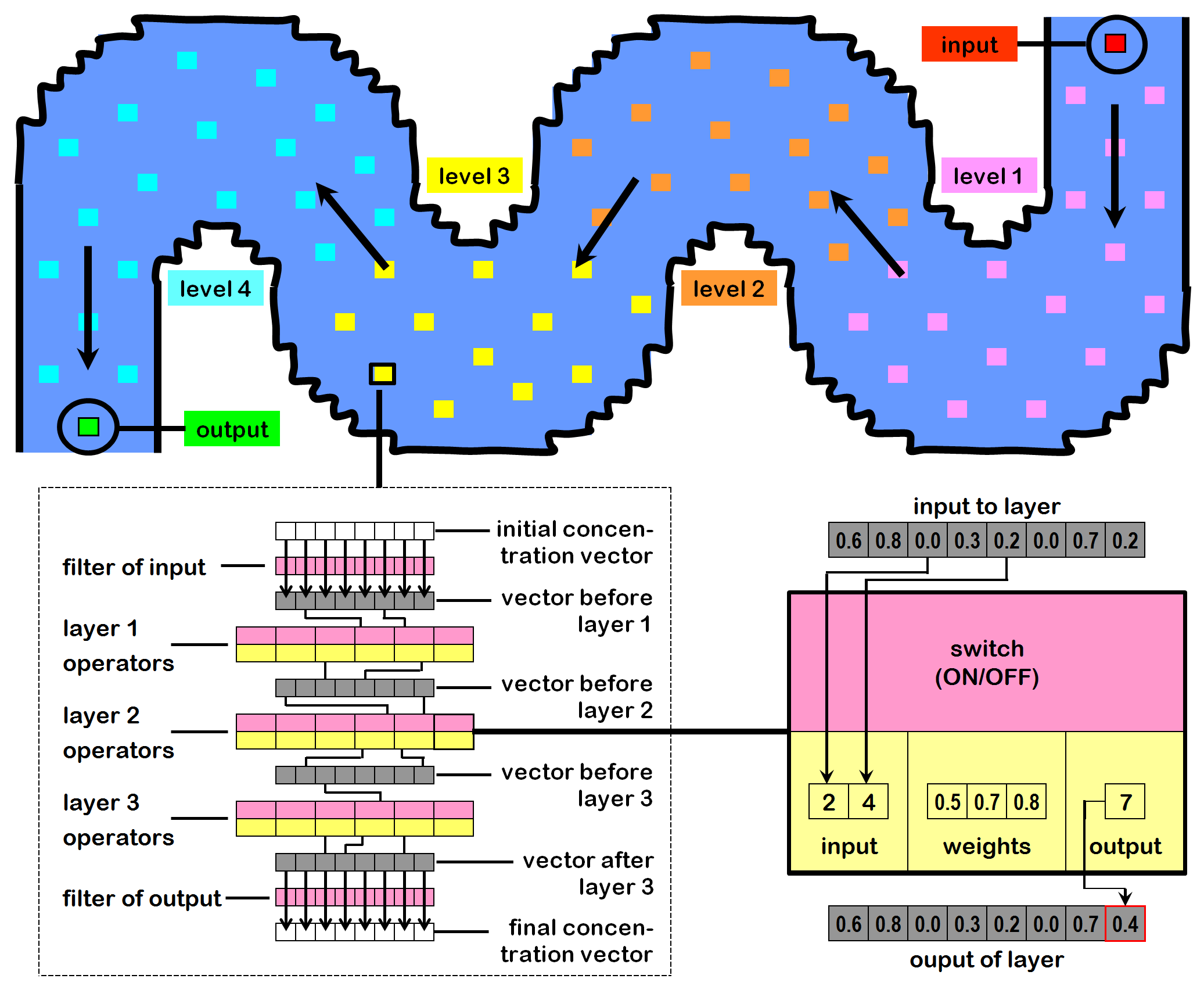}
\caption{Metabolic computation across multiple organisational levels. The minimal computational unit is the metabolic operator, encoded by a single metabolic gene. Metabolic operators, arranged in layers, build a feed-forward strictly-layered network which represent the processing device at the cellular level. Finally, cells across the shape, organised in layers, turn the entire organism into a computational device.}
\label{metabolic}
\end{center} \end{figure}

\section{Processing information in 3-dimensional structures: metabolic computation}

We have previously shown how ET can evolve 3D structures which can compute a predefined output for a given input \cite{Fontana10} using a new paradigm of computation, which we named ''metabolic computation''. To perform this computation,  each cell of the structure is associated with a \emph{concentration vector}, which contains the values of concentration for a number $S$ of substances ($S$=8 in the experiments described here). The input and the ouput are represented by vectors associated to two cells in a large multicellular structure. 


Metabolic computation occurs across multiple organisational levels (Fig.~\ref{metabolic}). The minimal computational unit is the \emph{metabolic operator}. Metabolic operators, arranged in 3 layers, build a network which represent the processing device at the cellular level. Finally, cells -- organised in levels -- compose a structure-wide computational network. In the present implementation, the computation is restricted to driver cells only (mainly for reasons of computational efficiency).

Each operator is encoded in a single metabolic gene, and has a structure composed of different fields. One field, called \emph{switch}, is a binary flag specifying whether the gene is active or not. Inactive genes are excluded from the computation. Other two fields, called \emph{operator input}, specify which positions of the concentration vector (i.e. which substances) are taken as the operator's inputs. Another field, \emph{operator output}, specifies which position of the concentration vector is affected by the operator's output. Finally, there is a field which contains a set of \emph{weights}.

The operator's output is given by the formula: $y = \sigma(\sum_{i}(weight(i)*ConcVector(input(i))+weight(0))$ ($y$ is a signed real number, which is added to the position of the concentration vector specified by the value of the operator output). The effect of various operators in the same layer on the concentration vector are additive. Two vectors of the same size as the concentration vector, called \emph{filters}, are dedicated to managing the exchange of substances between the cell and the external environment: if the filter position corresponding to a given substance is 1, the substance is allowed to enter or exit the cell (if it is 0 the substance is blocked). The computation is carried out for all layer 1 operators, then for all layer 2 operators, etc., until a final concentration vector is produced.

While operators are encoded in metabolic genes, and as such are common to all cells of the structure, we allow each cell to differentiate in terms of which operators can be used by each cell. This is specified in another field in the right part of developmental genes (not shown in Fig.~\ref{chevents}), which encodes changes to the switch of a number of metabolic genes. In this way, the repertoire of the cell's operators available for computation can become different in different cells, allowing for specialisation.


\begin{figure}[t] \begin{center}
\includegraphics[width=11.50cm]{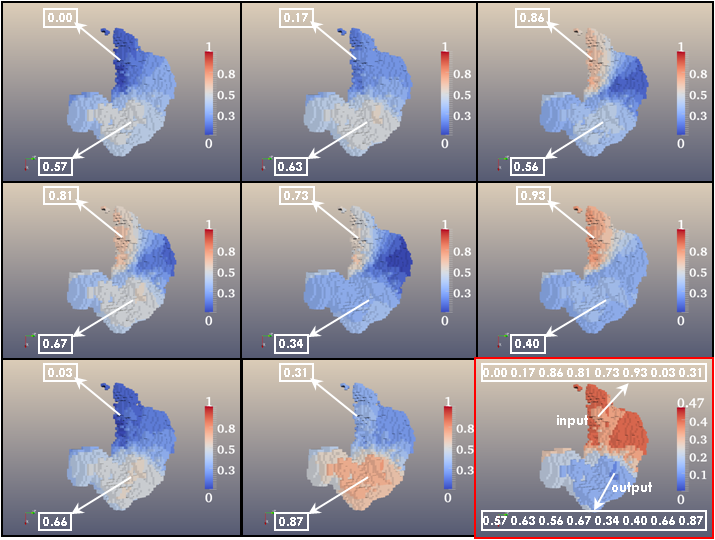}
\caption{Metabolic computation using a 3D multicellular structure. The first 8 panels (with black frames) show the concentration maps of each substance on the best individual evolved for 4 examples, using one of the examples. In the bottom-right panel (with a red frame) the colour of the cell corresponds to the normalised Euclidean distance of the input of this cell (a vector in 8-dimensional space of chemical concentrations) to the target output.}
\label{subsmap}
\end{center} \end{figure}

\begin{figure}[t] \begin{center}
\includegraphics[width=08.00cm]{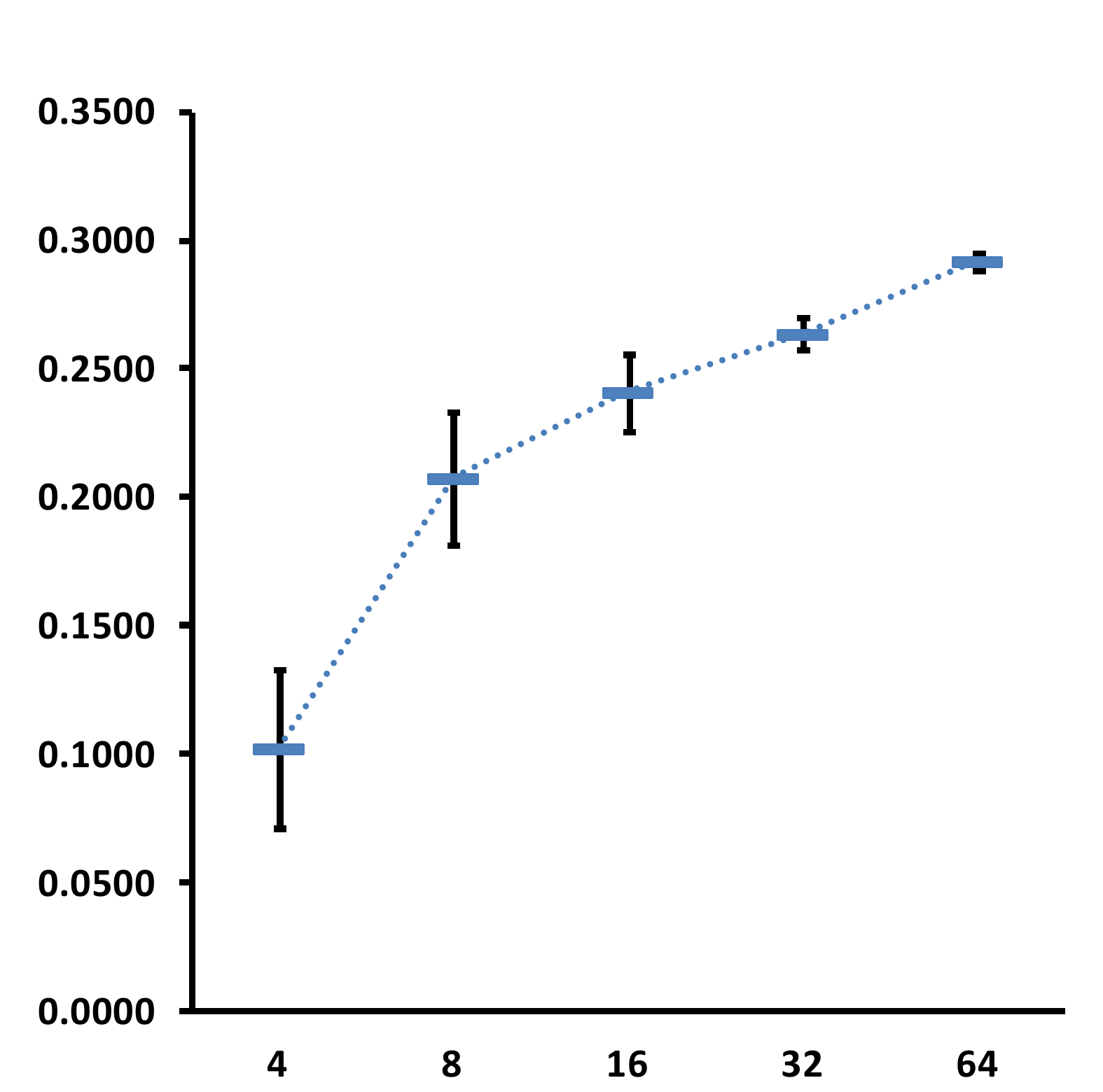}
\caption{Memory capacity in metabolic computation. The average normalised Euclidean distance between the output and the target concentration vectors is shown for 5 series of 10 independent evolutionary runs with the increasing number of input-ouput vectors (examples). The error bars correspond to standard deviations.}
\label{graph}
\end{center} \end{figure}    

\begin{figure}[t] \begin{center}
\includegraphics[width=11.50cm]{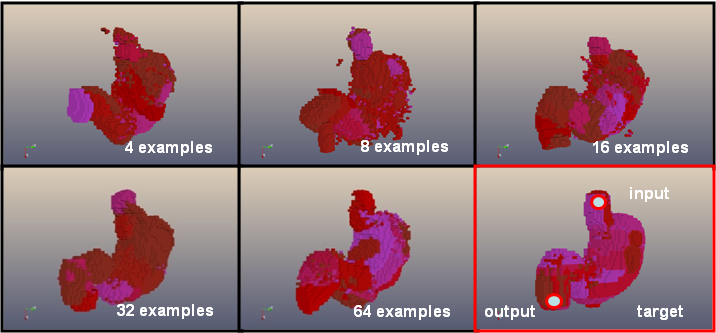}
\caption{Final stage of development for a 3D structure with the shape of a human stomach, able to perform metabolic computation. The first 5 panels (with black frames) represent the best individuals obtained in 10 independent runs with 4, 8, 16, 32 and 64 8-dimensional input-output pairs (examples), respectively. The bottom-right panel (with a red frame) represents the target shape; the circles indicate two points in space used to choose the closest driver cells taken as input and output.}
\label{stomachb}
\end{center} \end{figure}

The input and output cells are determined by choosing the two driver cells closest to two pre-defined points in the structure (Fig.~\ref{metabolic}). The remaining driver cells are assigned a \emph{level number} each. The value of this number depends on the distance of the cell from the input cell and the output cell: cells farther from the input and closer to the output receive a higher number. This converts the embryo's set of driver cells into a structure-wide strictly-layered feed-forward network. 

The processing is performed at the final stage of development (when the structure has reached its final shape) and starts by providing the input cell (level 0) with the input concentration vector. The concentration vector is processed by the cell's operators and transformed into a new concentration vector, which is then propagated to all level 1 cells, and so on. This simulates the diffusion of chemicals from one cell to other cells, and the substances are propagated to the next levels until the output cell (the last level) is reached. 

\section{Results and Discussion}

In the experiments in this paper, we have used a genetic algorithm to co-evolve shape and metabolism at the same time. In a previous paper \cite{Fontana10} we showed that co-evolution of shape and computation is possible using ET. The fitness function rewards both the adherence of the final shape to a predefined target shape and the adherence of the metabolic output to a given computational target (each fitness component has a 50\% weight). To compute the metabolic component of the fitness, the metabolic computation described for a number of examples (in this paper: 4, 8, 16, 32, or 64; each example corresponds to one input-output pair). The metabolic fitness is then calculated subtracting from 1 the Euclidean distance of the output from the target, averaged over all substances and examples.

For our genetic algorithm we have used constant population size (124 individuals), 50\% crossover probability, and mutation rate of 0.5\% per base in the genome. The performance of the genetic algorithm is boosted by a procedure called \emph{Germline Penetration} \cite{Fontana12}, which consists in creating new mobile sequences, able to match mobile codes generated during development, and copying them randomly into the left part of some developmental genes. In this way evolution is provided with ``good'' mobile sequences and becomes much faster.

In this work we have evaluated the memory capacity of the structures generated with ET. In each experiment the input cell was presented with a number of examples -- concentration vectors, each composed of 8 substances -- and the output cell was expected to match a concentration vector specific for a particular example. Both the input and the output vector were constructed by drawing random numbers from a uniform (0,1) distribution. We have run 5 series of evolutionary runs (with 10 independent runs each), one series for 4 examples, one for 8, and so on, up to 64. The shape component of the fitness rewarded the proximity of the final shape to a 3D structure composed of 20000 cells, representing a human stomach.
 
Our results show that the evolved metabolic networks gradually transform the input concentration vector into the output concentration vector (Fig.~\ref{subsmap}). The average normalised distance of two random 8-dimensional points whose coordinates are drawn from the uniform (0,1) distribution is 0.398866681..., so even the structures trained on 64 examples perform better than a completely random approximator (Fig.~\ref{graph}), while the variance of the approximation decreases with the increasing number of examples. 

As the memory task becomes more difficult with the increasing number of examples, the shape of the final structure of the champions in each series (Fig.~\ref{stomachb}) gets closer to the target. This better shape approximation indicates a trade-off between the part of the fitness function that rewards the shape and the part that accounts for computation: as it becomes more difficult to obtain a good result for the metabolic component of the fitness, more can be gained in evolutionary terms by optimising the shape.

The results described here confirm that it is possible to use ET to evolve a multicellular structure with a complex shape and capable of computation. Our future work will be directed in several directions. First, we plan to work on an encoding specifying the connectivity between the computational units in the structure, to allow for recurrent connections. Second, we would like to allow also normal cells to take part in the computation, hoping that our system can cope with the increased computational burden. Third, we would like to investigate what other computational tasks could be performed by complex structures obtained using ET.




\section*{Acknowledgements}

This work was supported by the Polish Ministry of Science and Higher Education (project 2011/03/B /ST6/00399); computational resources were provided by the Tri-city Academic Computer Centre (TASK) and the Interdisciplinary Centre for Molecular and Mathematical Modeling (ICM, University of Warsaw; project G33-8). We are grateful to Enrico Laeng for suggesting us a formula to calculate the expected performance of a random n-dimensional vector approximator. 

\bibliographystyle{eptcs}
\bibliography{ldanxwivac}

\end{document}